\documentclass{aa}  

\usepackage{graphicx}
\usepackage[dvipsnames]{xcolor}

\usepackage[varg]{txfonts}

\begin{document}
\title{Estimating the EOS from the measurement of NS radii with 5\% accuracy}
\authorrunning{Sieniawska et al.}
\titlerunning{NS EOS from 5\% accuracy radius measurement}

\author{M. Sieniawska$^a$, M. Bejger$^{a,b}$ and B. Haskell$^a$}
\institute{
$^a$ Nicolaus Copernicus Astronomical Center, Polish Academy of Sciences, ul. Bartycka 18, 00-716, Warsaw, Poland\\ 
$^b$ APC, AstroParticule et Cosmologie, Université Paris Diderot, CNRS/IN2P3, CEA/Irfu, Observatoire de Paris, Sorbonne Paris Cité, F-75205 Paris Cedex 13, France 
\\ \email{msieniawska@camk.edu.pl, bejger@camk.edu.pl, bhaskell@camk.edu.pl}
\\[3mm]
}

\date{Received ...; accepted ...}

\abstract
{Observations of heavy (${\simeq}2\,M_\odot$) neutron stars, such as PSR J1614-2230 and
PSR J0348+0432, in addition to the recent measurement of tidal deformability from the
binary neutron-star merger GW170817, place interesting constraints on theories of
dense matter. Currently-operating and future observatories, such as the Neutron
star Interior Composition Explorer (NICER) and the Advanced Telescope for High
ENergy Astrophysics (ATHENA) are expected to collect information on the
global parameters of neutron stars, namely masses and radii, with the accuracy of a
few percent. Such accuracy will allow for precise comparisons of measurements to models of compact objects and significantly improve our understanding of
the physics of dense matter.}
{The dense-matter equation of state is still largely unknown. Here we
investigate how the accuracy of the measurements expected from the NICER and ATHENA
missions will improve our understanding of the dense-matter interior of neutron stars.}
{We compare global parameters of stellar configurations obtained using three
different equations of state: a reference (SLy4 EOS) and two piecewise
polytropes manufactured to produce mass-radius relations indistinguishable from
the observational point of view i.e. within the predicted error of the radius
measurement. We assume observational errors on the radius determination
corresponding to the accuracies expected for the NICER and ATHENA missions. The effect 
of rotation is examined using high-precision numerical relativity computations. 
Due to the fact that masses and rotational frequencies might be determined very
precisely in the most optimistic scenario, only the influence of observational
errors on the radius measurements is investigated.}
{We show that ${\pm}5\%$ errors in radius measurement lead to ${\sim}10\%$ and ${\sim}40\%$ accuracy in central parameter estimation, for low-mass and high-mass neutron stars, respectively. Global parameters, such as oblateness and surface area, can be established with $8-10\%$ accuracy, even if only compactness (instead of mass and radius) is measured. We also report on the range of tidal deformabilities corresponding to the estimated masses of GW170817, for the assumed uncertainty in radius.} 
{}

\maketitle
\section{Introduction}
\label{sect:intro} 

Neutron stars (NS) are the most extreme material objects in the Universe.
Formed in the aftermath of core-collapse supernovae, some few thousands are
currently known. Most of them are seen as pulsars. These rotating compact
objects allow astronomers to perform direct measurements of global parameters of NSs and, indirectly, to investigate the properties of dense matter
in their interiors, which is in a state impossible to reproduce in terrestrial 
conditions. High-precision observations compared with theoretical models 
for the equation of state (EOS) of dense matter in NSs are 
currently the only way to study physics in such extreme conditions. 

Precise determinations of the masses of PSR J1614-2230 \citep{Demorest2010,
Fonseca2016,Arzoumanian2018} and PSR J0348+0432 \citep{Antoniadis2013}
have set an observational bound on the maximum mass of a NS not lower than about
$2\,M_\odot$ and constrain theoretical models of the EOS. Present
determinations of NS radii $R$ are roughly between 10-15 km (for a review see, e.g.
\citealt{Ozel2016,Fortin2016}), but due to systematic errors, uncertainties on the values are still large (see, e.g.
\citealt{Heinke2014,Fortin2015,EH16,Miller2016,Haensel2016}). Additionally,
GW170817, the recent first direct detection of gravitational waves from the
last orbits of a relativistic binary NS system \citep{Abbott2017a} yielded
tidal deformability parameter measurements that disfavour radii larger than 14
km at $1.4\,M_\odot$. 
 
Ongoing and future missions like the Neutron star Interior Composition Explorer
(NICER, \citealt{NICER}) and Advanced Telescope for High ENergy Astrophysics
(ATHENA, \citealt{Athena+}) are or will be able to measure NS
gravitational masses and radii, with accuracies of a few percent. 
Generally, the radius $R$ and the gravitational mass $M$ measurement methods consist of 
\begin{itemize}

\item Pulse profile modelling related to brightness variations from the
non-uniform surface of the NS. Such fluctuations may be caused by hot and
cold spots, originated by magnetic field in rotation-powered pulsars or by
non-uniform thermonuclear burning on the surface of an X-ray burster (see,
e.g., \citealt{Ozel2016b,Sotani2017}). According to \citet{Psaltis2014a}
and \citet{Psaltis2014b}, an uncertainty of $\lesssim 5\%$ in the NS radius
measurements is feasible, assuming that the rotational frequency of the object is
in the $300 - 800$ Hz range, and sufficiently long observations are possible
($10^{6}$ counts in the pulse profile). However, other authors pointed out that realistic observational errors might be larger, even up to $10\%$ (see e.g. \citealt{Lo2013, Miller2016}) and may depend strongly on the system geometry. Nevertheless, in our work we decided to assume $5\%$ accuracy, as an optimistic scenario. 

To break the degeneracy, four
quantities have to be measured - the amplitude of the bolometric flux
oscillation, the amplitude of its second harmonic, the amplitude of the
spectral colour oscillation and the phase difference between the bolometric
flux and colour oscillation. These requirements can be fulfilled by the NICER
mission's long exposure time and/or by combining pulse profile modelling methods
with other measurements (for example X-ray pulse modelling
from ATHENA and mass determination from radio timing). For lower spins
($\lesssim 300$ Hz), the amplitude of the second harmonic is too low to perform
a full analysis. In this case only measurements of the compactness $M/R$ are
possible. For much higher spins ($\gtrsim 800$ Hz), higher order multipoles
become important in the modelling and solutions of the field equations are
EOS-dependent. 

\item Observations of Eddington-limited X-ray bursts from accreting NSs, 
which may put constraints on the maximum radius of the compact object (see, e.g., 
\citealt{Galloway2008}).

\item Fitting spectra with an appropriate atmosphere model to observations
of the quiescent emission from Low-Mass X-ray Binaries. This method was proposed by
\citealt{Paradijs1979} and improved since that time by using realistic NS
atmospheres, as well as relativistic NS models and ray tracing  (e.g., \citealt{Vincent2017}). A recent analysis by
\citealt{Steiner2017} suggests NS radii between 10 to 14 km.

\item Detections of gravitational waves from NS binary systems, allowing for measurements 
of NS masses and radii from both observations of the late inspiral 
\citep{Bejger2005,DamourNV2012,Abbott2017a} and post-merger  
(e.g., \citealt{BausweinSJ2015,Abbott2017b,Annala2017,Margalit2017,Bauswein2017,Rezzolla2018}).
\end{itemize}

All these methods are necessarily limited by their intrinsic measurement
errors. The current state of the art is such that in several cases mass can be measured with much smaller errors than radii. Here we focus on future measurements of radii,  and in particular on
how the planned radius measurement accuracy reflects on the ability to
discriminate similar $M(R)$ relations obtained using different NS interior
prescriptions (that is, based on different EOSs).  Specifically, we
manufacture stable $M(R)$ sequences for parametric (piecewise-polytropic) EOSs and compare them to a reference sequence of configurations based
on the SLy4 EOS \citep{Douchin2001}. In this test case we investigate ways of
telling apart the $M(R)$ relations that are indistinguishable because of
observational errors in the radius measurement, which we assume to be equal to
5\% of the radius of the reference configuration based on the SLy4 EOS (the
$M(R)$ sequences of the piecewise-polytropes trace the $\Delta R={\pm}5\%$ outline of
the SLy4 EOS non-rotating $M(R)$ sequence). We also study different spin frequency cases: from
non-rotating objects to extremely rapidly spinning ones, in order to check if
rotation aids the discrimination between various functionals of the EOS
(mass, radius, quadrupole moment, moment of inertia). In addition, we study
cases in which the rotation rate is not known, and quantify the magnitude of
the errors on the EOS parameters related to the radius measurement error and/or the
lack of spin frequency measurement. 

The paper is composed as follows. In Sect.~\ref{sect:methods} we present
the EOS models, the numerical methods used to obtain the rigidly-rotating 
sequences of configurations, and list the global NS parameters of interest. 
Section \ref{sect:results} contains the results. Section \ref{sect:discussion} 
presents the discussion and conclusions.    

\section{Methods and equations of state}
\label{sect:methods}

The state of matter is relatively well known below the nuclear saturation
density $\rho_{s} = 2.7 \cdot 10^{14}$ g cm$^{-3}$. Above this density several
competing theories describing the EOS of dense matter exist (for a textbook
review, see e.g., \citealt{HaenselPY2007}). In order to compare the effects of
the uncertainty in the radius measurement on the EOS, we choose as the reference
the SLy4 EOS \citep{Douchin2001}, consistent with recent radius and mass
constraints, described in the introduction. Furthermore, we select two
parametric EOSs, named Model1 and Model2, manufactured to be barely consistent
with the expected radius uncertainty measurement with respect to the
non-rotating reference model.

\begin{figure}[ht]
\resizebox{\columnwidth}{!}
{\includegraphics{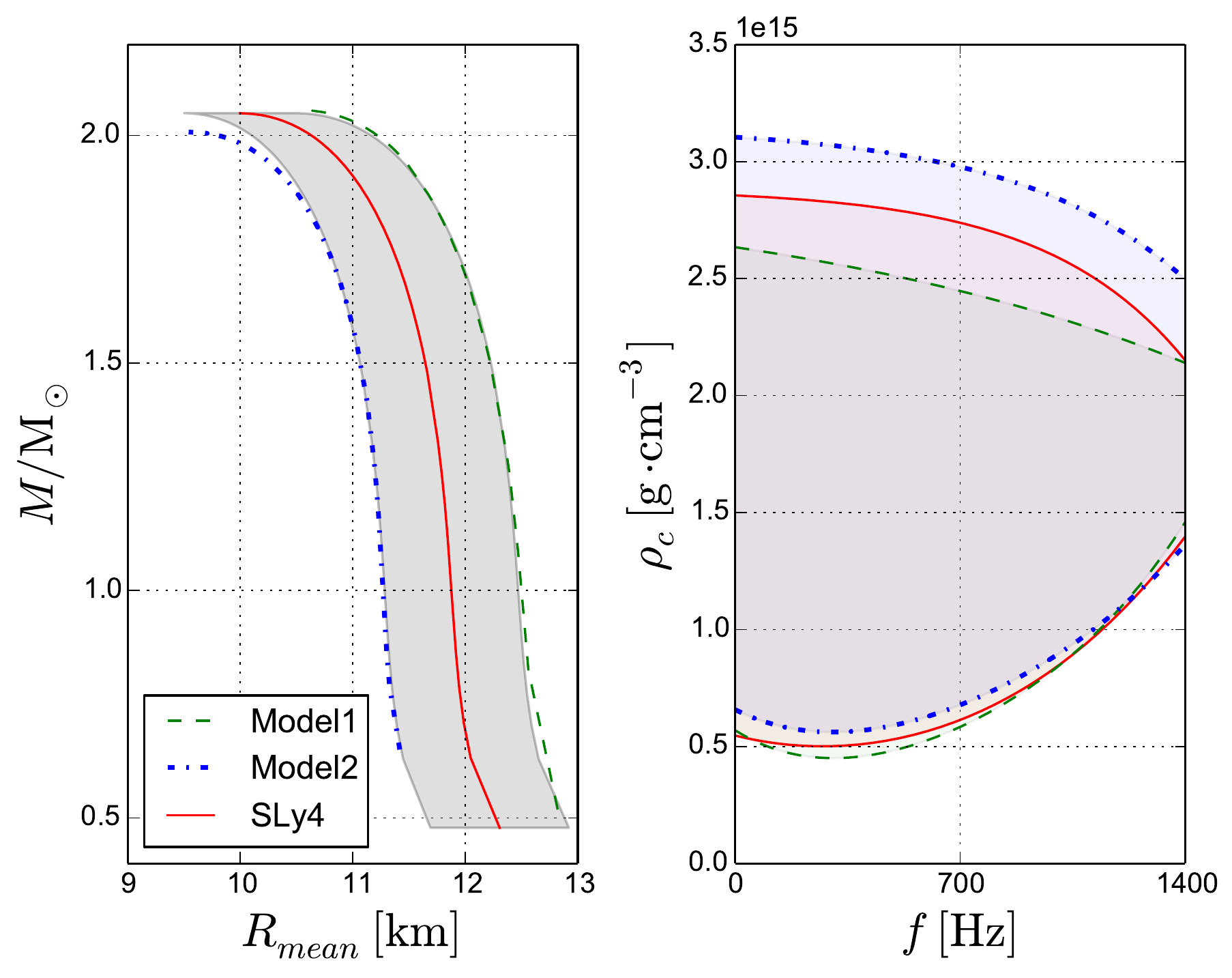}}
\caption{(Color online) Right panel: Range of central densities $\rho_{c}$ for stars rotating with the frequency $f$ up to 1400 Hz. Left panel: non-rotating configurations for the SLy4 EOS (red), 
Model1 EOS (green) and Model2 EOS (blue), following the $\Delta R = \pm 5\%$ uncertainty 
in the radius measurement (gray region).}
\label{fig:mrtov_rhof}
\end{figure}

Model1 and Model2 EOSs were constructed using the SLy4 prescription of the crust 
for densities lower than the nuclear saturation density, and with three piecewise relativistic 
polytropes 
\begin{equation}
P(n) = \kappa_{i}n^{\gamma_{i}},\quad \epsilon(n) = \rho c^2 = \frac{P}{\gamma_{i} - 1} + nm_{b_{i}}c^{2}, 
\label{eq:poly}
\end{equation}   
for higher densities, where $P(n)$ and $\epsilon(n)$ denote the pressure and
mass-energy density as function of the baryon density $n$, and $\kappa_{i}$,
$\gamma_{i}$ and $m_{b_{i}}$ are the pressure coefficient, the polytropic index
(characterizing the stiffness of the EOS at given density) and the baryon mass
for a given polytropic $i$-th segment ($i = 1,\dots,3$). Index $\gamma_{i}$ is
a parameter of choice, and $\kappa_{i}$ and $m_{b_{i}}$ are fixed for a
polytropic segment by the mechanical and chemical equilibrium.
 
\begin{table}[h]
\centering
\begin{tabular}{|c|c|c|}
\hline
par. & Model1 & Model2 \\
\hline
$\gamma_{1}$ & 3.20 & 2.50 \\
\hline
$\gamma_{2}$ & 2.83 & 3.22 \\
\hline
$\gamma_{3}$ & 2.50 & 3.00 \\ 
\hline
$n_{b, 12}$ & 0.21 & 0.24 \\
\hline
$n_{b, 23}$ & 0.70 & 0.50 \\
\hline
$m_{b,1}$ & 1.017982 & 1.016573 \\
\hline
$m_{b,2}$ & 1.014858 & 1.021916 \\
\hline
$m_{b,3}$ & 0.977851 & 1.015670 \\
\hline
$\kappa_{1}$ & 0.006646 & 0.006646 \\
\hline
$\kappa_{2}$ & 0.008745 & 0.003538 \\
\hline
$\kappa_{3}$ & 0.016621 & 0.005042 \\
\hline

\end{tabular}
\vskip 0.5em 
\caption{Parameters of the three polytropes employed for the EOSs of Model1 and
Model2 (indices correspond to the number of the polytrope). $\gamma$ is the index of polytrope, $n_{b}$ is
baryon density, where change of the polytrope occurred, $\kappa$ is the pressure coefficient and $m_{b}$ the baryon mass at the point where the crust joins with the first polytrope ($m_{b,1}$) or where two polytropes join ($m_{b,2}$ and $m_{b,3}$). Other
parameters were calculated from the conditions of mechanical and chemical
equilibrium.}
\label{tab:poly_parameters}
\end{table}

In this study we limit ourself to stationary, axisymmetric, rigidly-rotating NS
configurations.  Non-rotating static NS solutions are obtained by solving the
TOV equations \citep{Tolman1939,OppenheimerV1939}. Sequences of rotating stars
parametrized by the spin frequency $f$ and the EOS parameter at the stellar
center (e.g., the central pressure $P_c$) are obtained by solving coupled
partial differential equations using a multi-domain spectral methods library
{\tt LORENE}\footnote{\tt http://www.lorene.obspm.fr} (Langage Objet pour la
RElativité NumériquE, \citealt{Gourgoulhon2016}) {\tt nrotstar} code
(\citealt{BonazzolaGSM1993}, \citealt{Gourgoulhon1999},
\citealt{Gourgoulhon2010}). The accuracy of the solutions is inspected by
checking the validity of the 2D general-relativistic virial theorem
\citep{BonazzolaG1994}. Global parameters describing the NS are gravitational
and baryon masses $M$ and $M_b$, angular momentum $J$ and quadrupole moment
$Q$; their definition can be found in \citet{BonazzolaGSM1993,Gourgoulhon2010}.
From electromagnetic observations one obtains some estimation of the flux from
the stellar surface, $F_{\infty} \propto T^{4} R^2$, proportional to its
effective temperature $T$ and the size $R$ of the star. For a rotating star its
visible size and radius are not uniquely defined (depends on many factors:
viewing angle, compactness, rotation rate, physical parameters of the
atmosphere, see e.g., \citealt{Vincent2017} for details); in order to compare
configurations rotating with different rates we adopt the mean radius $R_{mean}
= \sqrt{S/4\pi}$, where $S$ is the surface area of the star, as a sufficiently good
approximation. 

We also compare the tidal deformabilities related to different models. The
tidal deformability $\lambda$ which represent the reaction of the star on the
external tidal field (such as that in a tight binary system) were obtained in
the lowest-order approximation by integrating the TOV equations supplemented by
an additional equation for the second tidal Love number $k_2$
\citep{FlanaganH2008,VanOeverenF2017}, $\lambda = 2R^5k_2/3$, where $k_2$ is
the quadrupole Love number \citep{Love1911} and $R$ is the non-rotating star
radius.  We use the normalized value of the $\lambda$ parameter, $\Lambda =
G\lambda\left( GM/c^2\right)^{-5}$.

\section{Results}
\label{sect:results}

Parametric models, denoted Model1 and Model2, are chosen in such a way to
produce non-rotating $M(R)$ relations tracing the $\Delta R = \pm 5\%$ outline
of the SLy4 EOS non-rotating $M(R)$ sequence; they are plotted in the left
panel in Fig.~\ref{fig:mrtov_rhof}. At the present moment the data-analysis
methods allow to treat objects with rotational frequency $\lesssim 800$ Hz, as
was mentioned in the Sect.~\ref{sect:intro}. Nevertheless, one can expect that
future theoretical and observational progress will enable studies of NSs with
larger spins. We therefore research a broader range of spin frequencies,
$0-1400$ Hz, as shown in the right panel in Fig.~\ref{fig:mrtov_rhof}. The EOSs
parameters are collected in Table~\ref{tab:poly_parameters}. In addition to the
required difference in radius, we select the parameters so that (i) the maximum
mass $M_{max} > 2M_{\odot}$, (ii) the speed of sound $\sqrt{{\partial
P}/{\partial\rho}}$, is always smaller than the speed of light $c$ for stable
configurations. 

\subsection{Non-rotating neutron stars}
\label{sect:nonrotNS}

\begin{figure}[ht]
\resizebox{\columnwidth}{!}
{\includegraphics{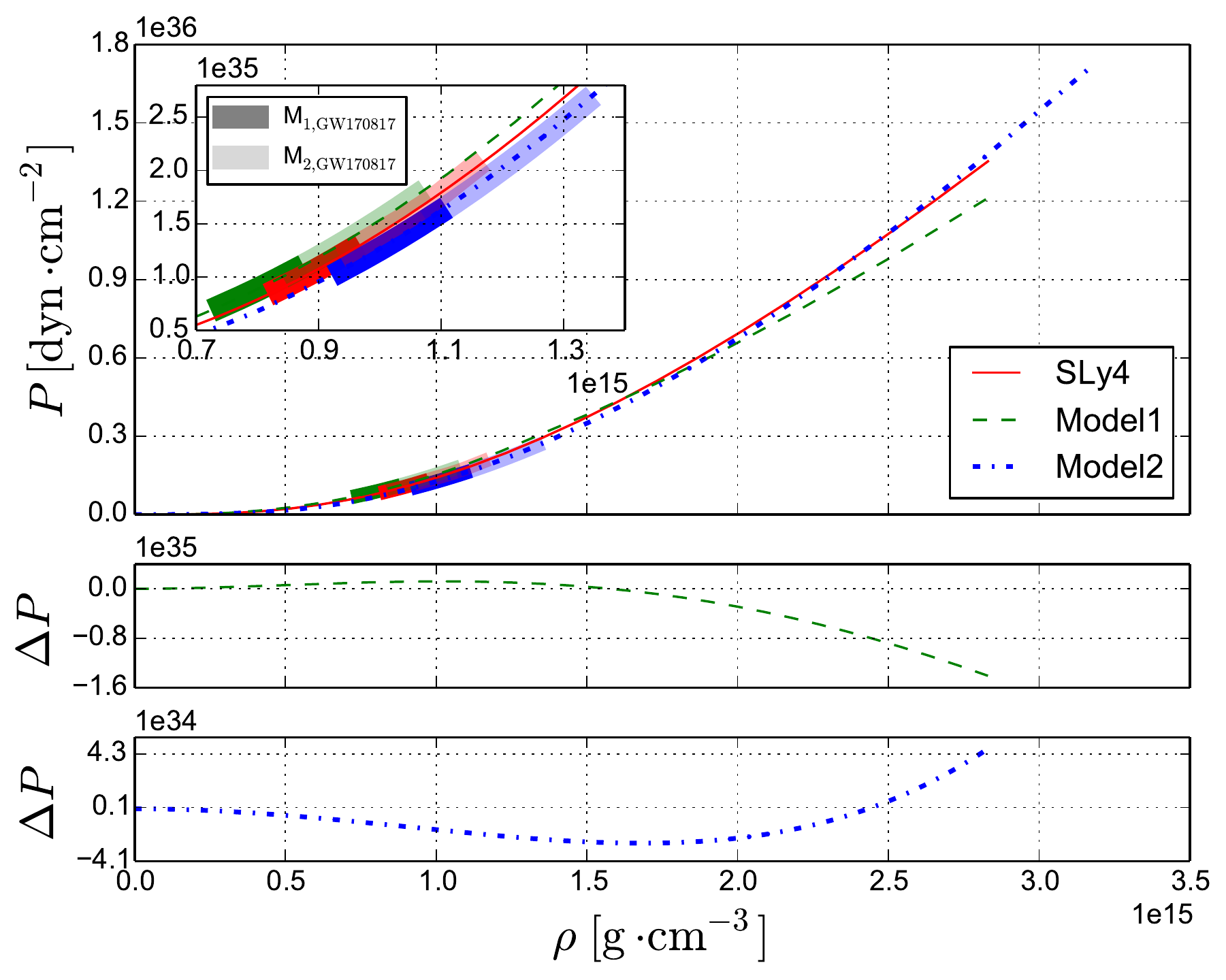}}
\caption{(Color online) Upper panel: $P(\rho)$ profiles corresponding to the
$M(R)$ relations from the left panel of Fig.\;\ref{fig:mrtov_rhof}, with the
SLy4 model denoted by a red solid line, Model1 by a green dashed line and
Model2 by a blue dash-dotted line. Thick semi-transparent lines denote
$P(\rho)$ ranges for the two components of the GW170817 binary neutron star
system (\citealt{Abbott2017a}, with mass estimates using the low-spin priors).
We also plot the pressure difference between the SLy4 model and Model1 (dashed
green line, middle panel) and Model2 (dash-dotted blue line, lower panel).}
\label{fig:profiles}
\end{figure}

The pressure-density $P(\rho)$ relations for non-rotating configurations are presented in Fig.
\;\ref{fig:profiles}. The two bottom panels show the differences between the reference SLy4 
model, Model1 (bottom panel) and Model2 (middle panel). Note that Model1 is initially 
stiffer than the SLy4 EOS, which provides a larger radius. For higher densities Model2 
becomes stiffer, which is necessary to reach the desired maximum mass. 

\subsection{Slowly-spinning neutron stars}
\label{sect:slowrotNS}

As was mentioned in the Sect.~\ref{sect:intro}, a full pulse profile analysis
cannot be performed for NSs with rotational frequencies $\lesssim 300$ Hz,
because the signal cannot be distinguished from a sinusoid and its second
harmonic is too weak to perform a Fourier decomposition. As a consequence, only
compactness can be measured (here defined as $\mathcal{C}= 2GM/R_{mean}c^{2}$
ratio, where $G$ is the gravitational constant, and $c$ is the speed of light).
Here we analyse the influence of the measured compactness on global and central
NS parameters.

\begin{figure}[ht]
\resizebox{\columnwidth}{!}
{\includegraphics{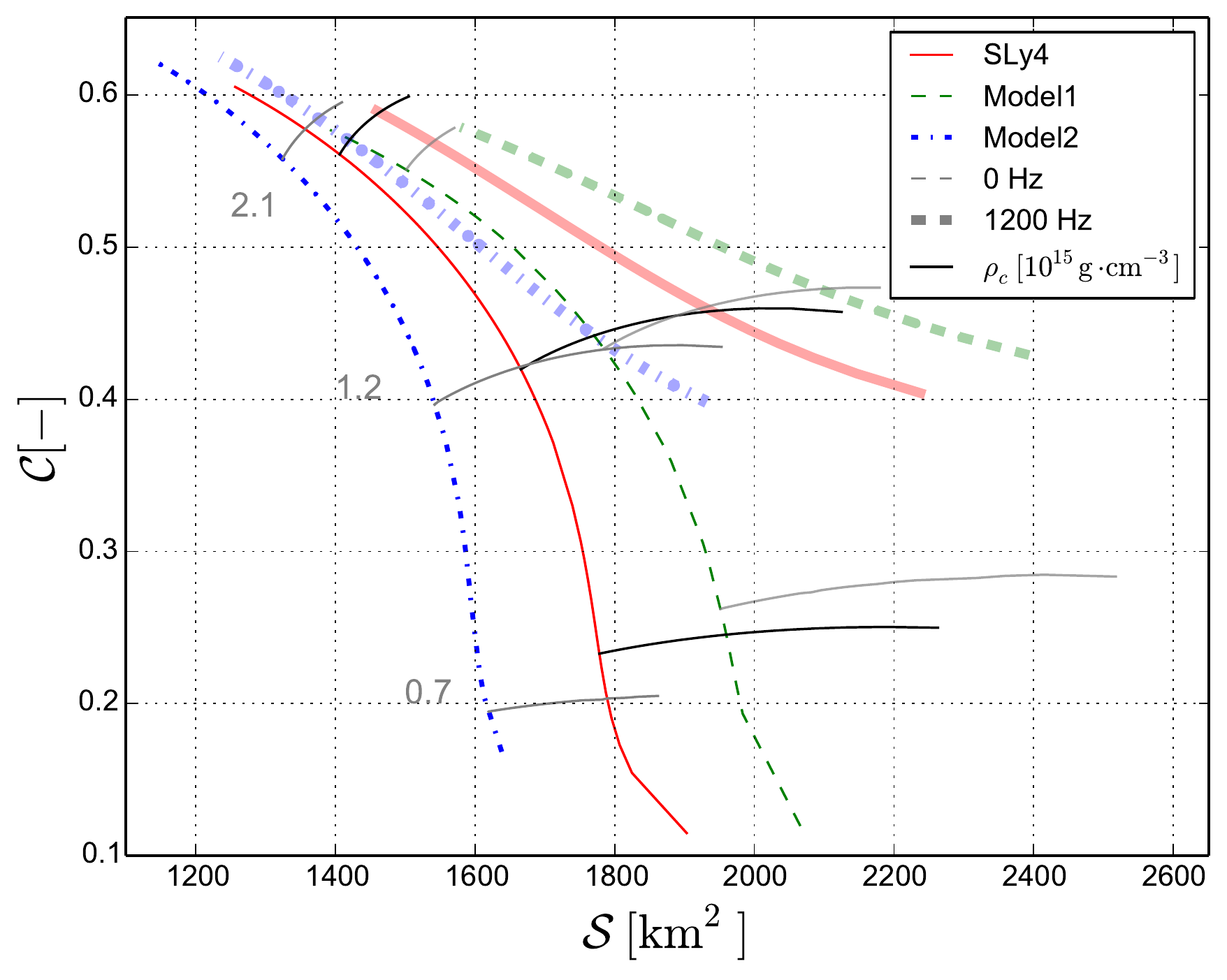}}
\caption{(Colour online) Compactness versus surface area for three EOSs: SLy4 (solid red), Model1 (dashed green) and Model2 (dash-dotted blue), for two rotational frequencies: 0 Hz (thin lines) and 1200 Hz (thick lines). Lines of constant central density are marked in grey (dark grey for SLy4, medium grey for Model2 and light grey for Model1). Corresponding values of the central density are given on the left side of the plot (in units of $10^{15}\ \mathrm{g \cdot cm}^{-3}$).}
\label{fig:SCR}
\end{figure}

In Fig.\;\ref{fig:SCR} we show the compactness as a function of the NS surface area.
Selected values of constant $\rho_{c}$ are also marked. Configurations with
$\mathcal{C} \gtrsim 0.6$ are possible only for Model2. Additionally, if in
addition to $\mathcal{C}$ the rotational frequency $f$ is also known, the NS
surface area can be estimated with $\sim 10\%$ accuracy, for all values of
$\mathcal{C}$ and $f$. Unfortunately, obtaining the central parameters, such as
the central density $\rho_{c}$, is difficult when only the compactness, instead
of $M$ and $R$, is known. For example, a central density $\rho_{c} = 0.7 \cdot
10^{15}\mathrm{g \cdot cm}^{-3}$ spans $\mathcal{C}$ between $0.2$ and $0.3$. 

\subsection{Unknown rotational frequency}
\label{sect:unknownrotNS}

\begin{figure}[ht]
\resizebox{\columnwidth}{!}
{\includegraphics{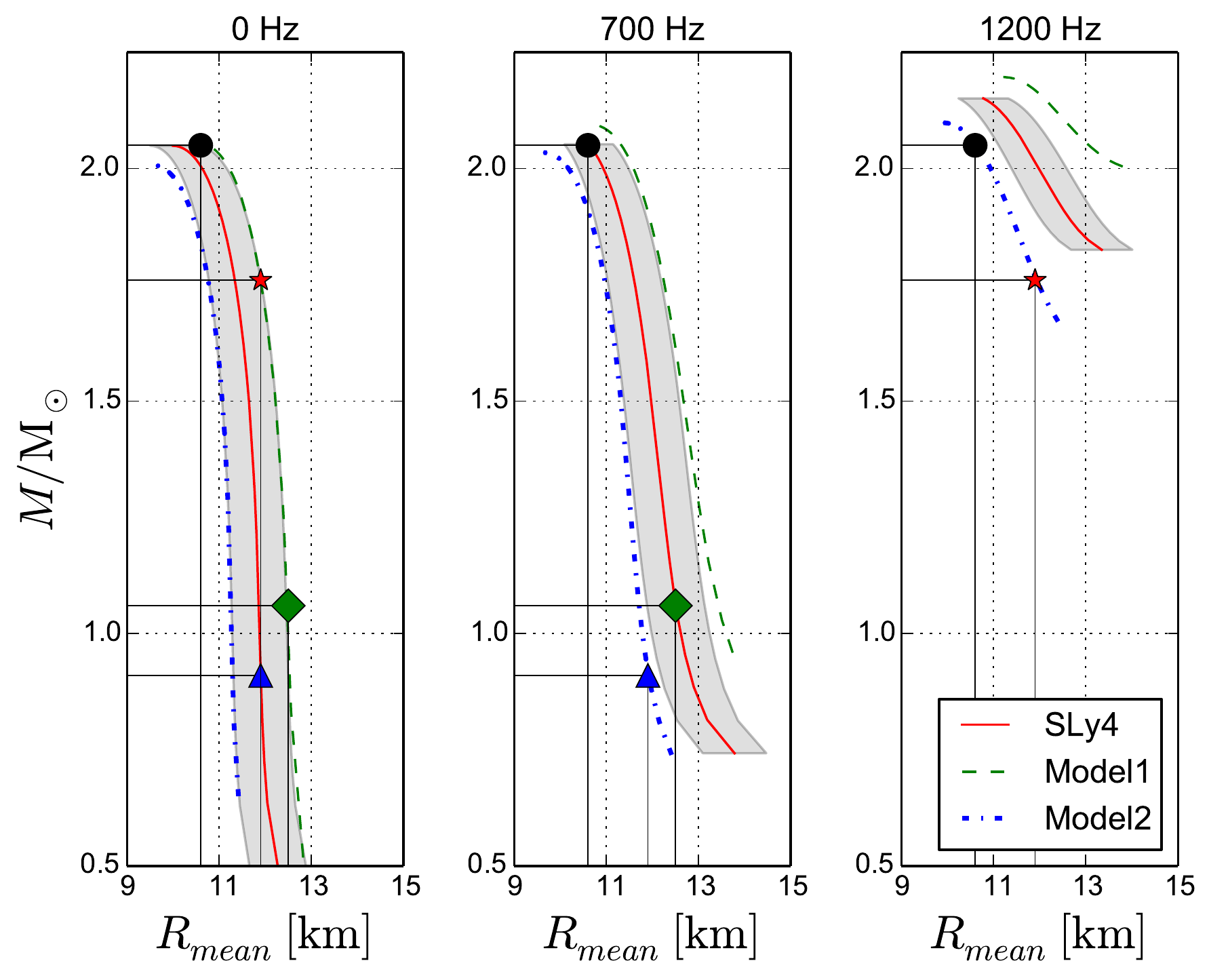}}
\caption{(Color online) Stable mass-radius relations for three models: SLy4
model (red solid line), Model1 (green dashed line), Model2 (blue dash-dotted
line) and three spin frequencies: $0$ Hz (left panel), $700$ Hz (middle panel)
and $1200$ Hz (right panel). Shaded regions correspond to $\Delta R = \pm 5\%$ of radii
measurements for the non-rotating SLy4 model. Note that some lines of stable configurations with different EOS
and $f$, but similar masses and radii can cross each other. Such cases are denoted by markers: black dots correspond to $R_{mean} \approx 10.6$ km and $M \approx 2.05 M_{\odot}$, red stars to $R_{mean} \approx 11.9$ km and $M \approx 1.76 M_{\odot}$, green diamonds to $R_{mean} \approx 12.5$ km and $M \approx 1.06 M_{\odot}$ and blue triangles to $R_{mean} \approx 11.9$ km and $M \approx 0.91 M_{\odot}$. Models with a fixed mass and radius may thus belong to configurations with different EOS, depending on the value of the spin. This adds additional ambiguity to the determination of the EOS from mass and radius measurements, if the rotation rate of the star is not known.}
\label{fig:MR}
\end{figure}

We limit our analysis of rotating stars to axisymmetric, rigidly-rotating,
stable cases (frequencies below the Keplerian frequency). We have studied a wide range of possible spin frequencies and central parameters, as shown on the right panel on the Fig.~\ref{fig:mrtov_rhof}. The fastest-known
pulsar, PSR J1748-2446ad \citep{Hessels2006} rotates at $716$ Hz. However, to fully cover the possible parameter space, we survey spin frequencies much higher than this current limit, up to
$1200$ Hz, which is just above the spin frequency of $1122$ Hz, suggested for XTE J1739-285 (reported
by \citealt{Kaaret2007}, but not confirmed burst oscillation frequency).
The typical accuracy of the results, monitored by the GRV2 virial error, is of the
order of $10^{-6}$. Figure\;\ref{fig:MR} shows stable configurations for the
three models: no rotation (left panel), $700$ Hz (middle panel) and $1200$ Hz
(right panel). For the $700$ Hz spin frequency, the $M(R)$ relations for
Model1 and Model2 go beyond the shaded region representing the tentative
observational errors. The deviation is stronger for less-massive stars,
$\lesssim 1.4M_{\odot}$. As expected, for extremely rapidly-rotating NSs ($1200$
Hz) only objects with large masses, ${\approx}2 M_\odot$, are able to
counterbalance the centrifugal force; all stable Model1 and Model2
configurations are outside the $\Delta R = \pm 5\%$ region.  In the following, we will
discuss the scenario in which, when the measurements of spin are uncertain,
configurations of different EOS and spin may have the same mass and radius.
Such ambiguous configurations are marked with symbols on Fig.\;\ref{fig:MR}.  

\begin{figure}[ht]
\resizebox{\columnwidth}{!}
{\includegraphics{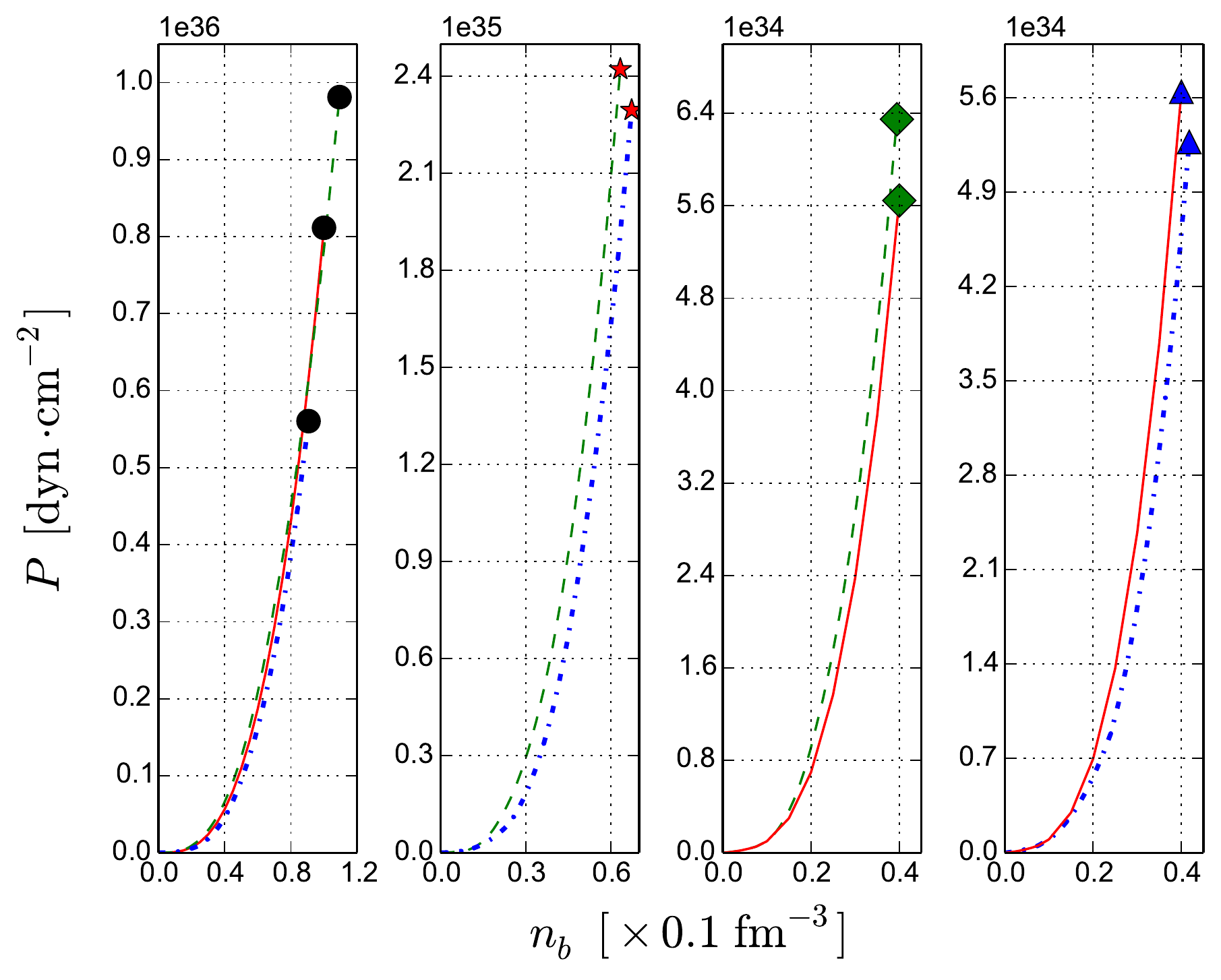}}
\caption{(Color online) $P(n_b)$ profiles for NSs marked on Fig.~\ref{fig:MR} with corresponding markers. Objects have the same $M$ and $R$, but different spin frequencies $f$ and EOSs.}
\label{fig:MRspecial_pres}
\end{figure}

For example, at $R_{mean} \approx 10.6$ km and $M \approx 2.05 M_{\odot}$
(denoted by a black dot), three different NSs are possible: a non-rotating
configuration described by Model1, the reference model SLy4 rotating at $f =
700$ Hz, and Model2 at $f=1200$. Such a set-up corresponds to a situation in
which the masses and radii are determined, but the spin frequency is unknown,
e.g., for a NS in a binary system for which bursts are observed, but which is not observed as a pulsar. We
compare these configurations in the $P(n_b)$ plane in Fig.~\ref{fig:MRspecial_pres}. 

This illustrates the importance of the spin information for the inference of the
EOS from observations. The most extreme example is shown in the left panel
of Fig.~\ref{fig:MRspecial_pres}: the accuracy of the estimation of the central
density between Model1 with $f = 0$ Hz, and Model2 with $f = 1200$ Hz is approximately
$40\%$. Additionally, the pressure difference between the Model1 ($f = 0$ Hz) and
SLy4 ($f = 700$ Hz) configurations is almost $20\%$. Other properties of the NSs from this
case are shown in Table \ref{tab_of_special}: the surface area $\mathcal{S}$,
oblateness $\mathcal{O}$, T/W ratio and the angular momentum $J$.
Interestingly, the configuration described by Model1 ($f = 0$ Hz) has almost
the same surface area $\mathcal{S}$ as the fast-spinning ($f = 1200$ Hz)
Model2 configuration. In the other panels of Fig.~\ref{fig:MRspecial_pres}, 
less extreme cases are presented; their differences between central pressures are around $10\%$. 

\begin{table}[t]
\begin{tabular}{|l|c|c|c|c|c|}
\hline
Model&$f$&$\mathcal{S}$&$\mathcal{O}$&T/W&J\\
~&[Hz]&[$10^{3}$ km$^{2}$]&&&[GM$_{\odot}^{2}$/c]\\
\hline 
SLy4&$700$&$1.4160$&$0.9463$&$0.0172$&$1.0552$\\ 
\hline
Model1&$0$&$1.4520$&$1.0000$&$0$&$0.0000$ \\
\hline
Model2&$1200$&$1.4958$&$0.8109$&$0.0614$&$1.8948$\\ 
\hline
\end{tabular}
\caption{Parameters (rotational frequency $f$, surface area $\mathcal{S}$, oblateness $\mathcal{O}$, T/W ratio and angular momentum $J$) of NSs marked in Fig.~\ref{fig:MR}, with the mean radius $R_{mean} \approx 10.6$ km and mass $M \approx 2.05 M_{\odot}$.}
\label{tab_of_special}
\end{table}

\begin{figure}[ht]
\resizebox{\columnwidth}{!}
{\includegraphics{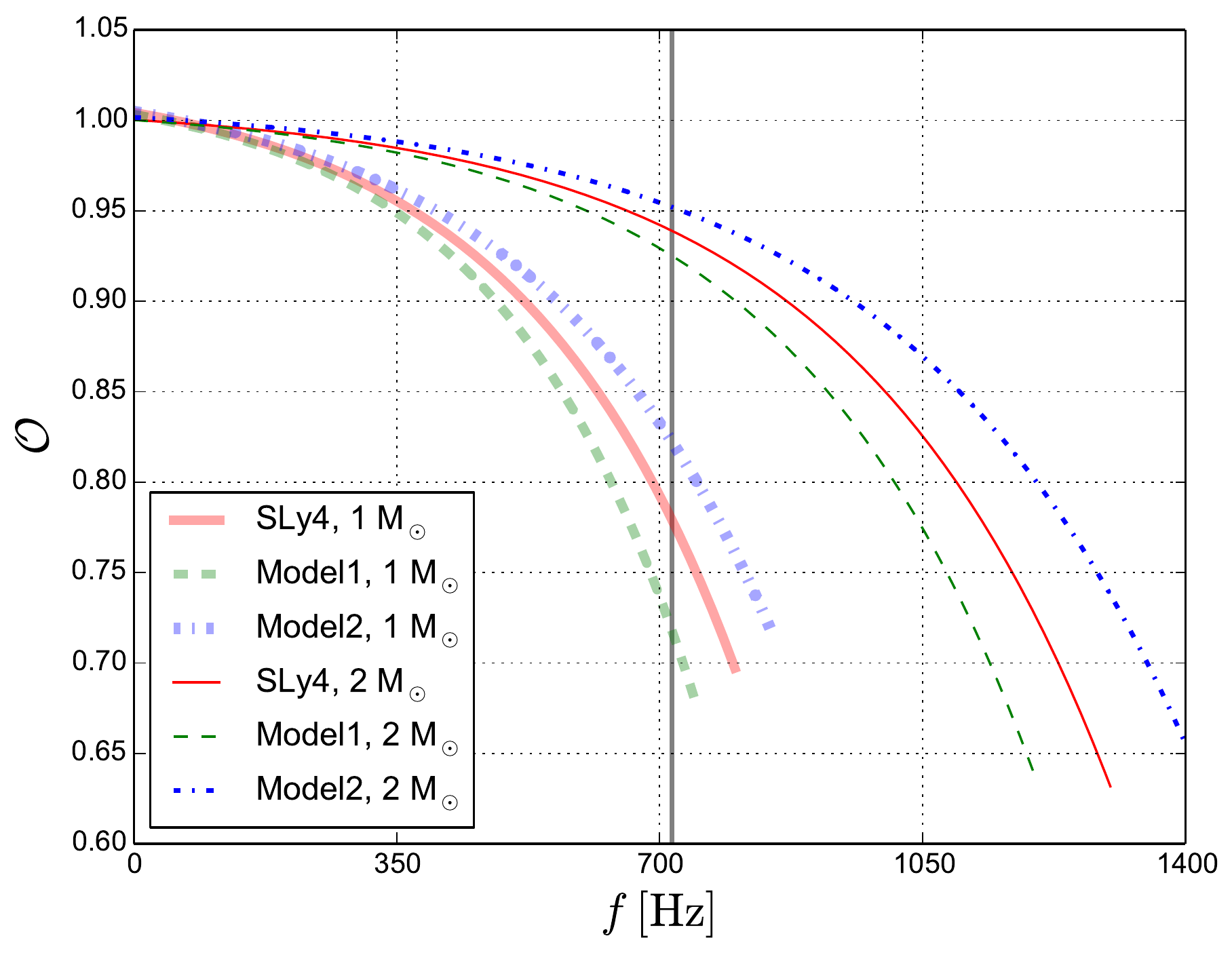}}
\caption{(Color online) Oblateness $\mathcal{O}$ versus rotational frequency for SLy4 model (solid red line), Model1 (dashed green line), Model2 (dash-dotted blue line) for $1 M_{\odot}$ and $2 M_{\odot}$ masses. The frequency of PSR J1748-2446ad ($716$ Hz) is marked as by the gray vertical line.}
\label{fig:oblateness}
\end{figure}

\begin{figure}[ht]
\resizebox{\columnwidth}{!}
{\includegraphics{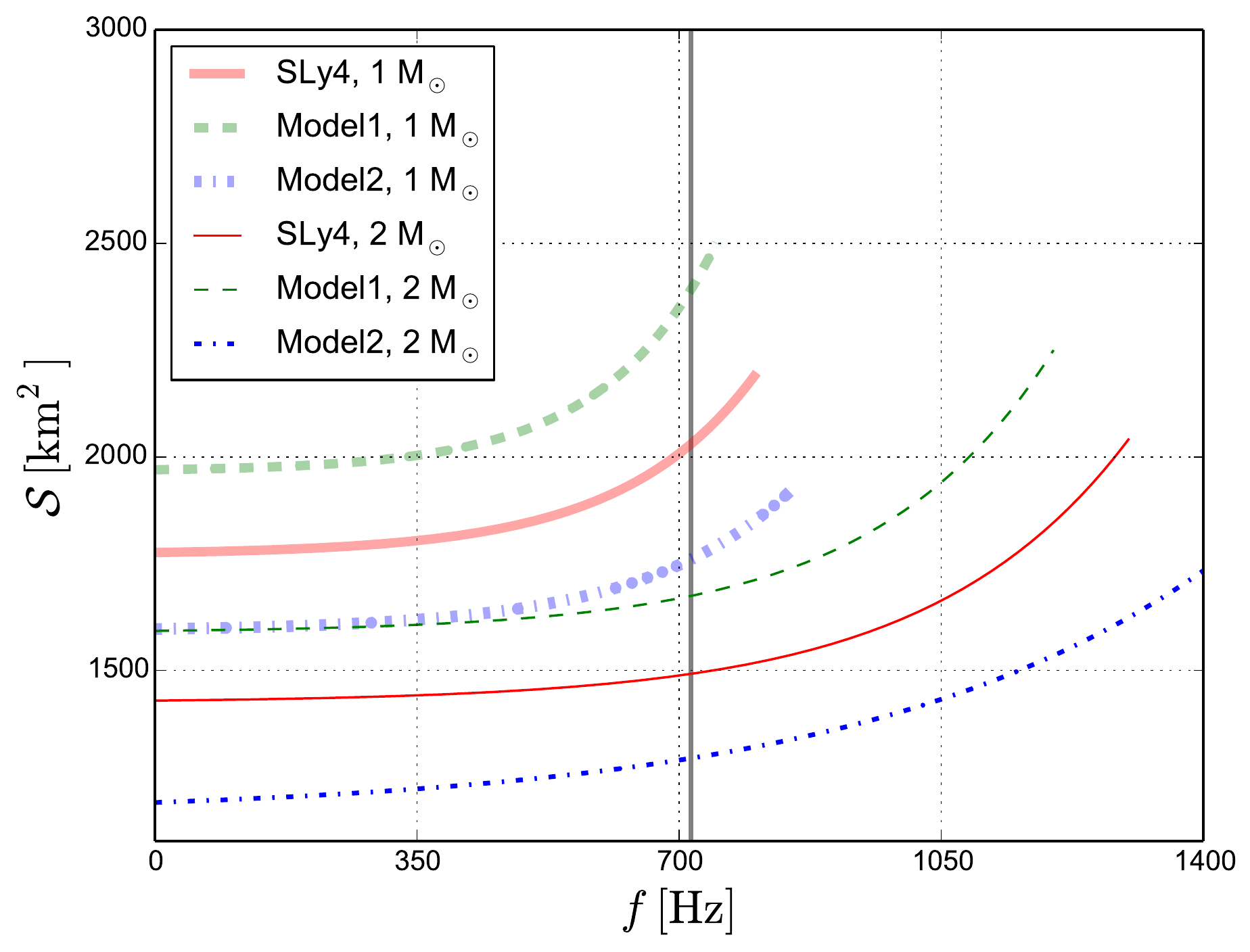}}
\caption{(Color online) Surface area $\mathcal{S}$ versus rotational frequency for the SLy4 model (solid red line), Model1 (dashed green line) and Model2 (dash-dotted blue line), for $1 M_{\odot}$ and $2 M_{\odot}$ masses.}
\label{fig:area}
\end{figure}

The influence of rotation on the global parameters, like the oblateness
$\mathcal{O}$ and the surface area $\mathcal{S}$, differs between low-mass and
massive NSs, as is shown in Figs.~\ref{fig:oblateness} and \;\ref{fig:area}.
Grey vertical lines correspond to PSR J1748-2446ad. As expected, from these
two figures one can notice that NSs with masses $\approx 1 M_{\odot}$ reach
the mass-shedding frequency much faster ($\approx 750$ Hz for Model1 and
$\approx 850$ Hz for Model2) than in the case of massive NSs. If PSR
J1748-2446ad is a light NS, its oblateness is between $\approx 0.73 - 0.83$. In
other words, a $\Delta R = \pm 5\%$ accuracy in the radius measurement leads to $\pm
8\%$ accuracy in $\mathcal{O}$ and to $\pm 10\%$ accuracy in $\mathcal{S}$
estimation.  For stars with $M \approx 2 M_{\odot}$, the dependence of
oblateness on EOSs is much weaker: at the spin frequency of PSR J1748-2446ad the
oblateness is around $0.95$ for all three models ($\pm 1\%$ accuracy). For the
faster rotation, $f \approx 1200$ Hz, the error of the $\mathcal{O}$ increases up
to $\pm 11\%$. 

The NS surface area for masses $M \approx 1 M_{\odot}$ is almost constant
with increasing rotational frequency. A significant increase of $\mathcal{S}$ appears for $f \gtrsim 650$ Hz, close to the mass-shedding limit. Massive NSs are
also not very susceptible to centrifugal force deformations. Significant
deformations occur at $\approx 1000$ Hz. For the whole range of $f$ (for NSs
with $M \approx 1 M_{\odot}$ and $M \approx 2 M_{\odot}$), $\mathcal{S}$
changes about $\pm 10\%$. For PSR J1748-2446ad, for which the mass is
unknown, the error in the surface area estimation is large: $\mathcal{S}$ for Model2 $2
M_{\odot}$ NS is twice as small as for Model1 $1 M_{\odot}$ NS. As was mentioned in 
Sect.~\ref{sect:slowrotNS}, a precise measurement of compactness $\mathcal{C}$ leads to a $\pm 10\%$ accuracy in the surface area estimation, if $f$ is known. This is comparable with the result that we get for low-mass NSs with unknown spins. 

\begin{figure}[ht]
\resizebox{\columnwidth}{!}
{\includegraphics{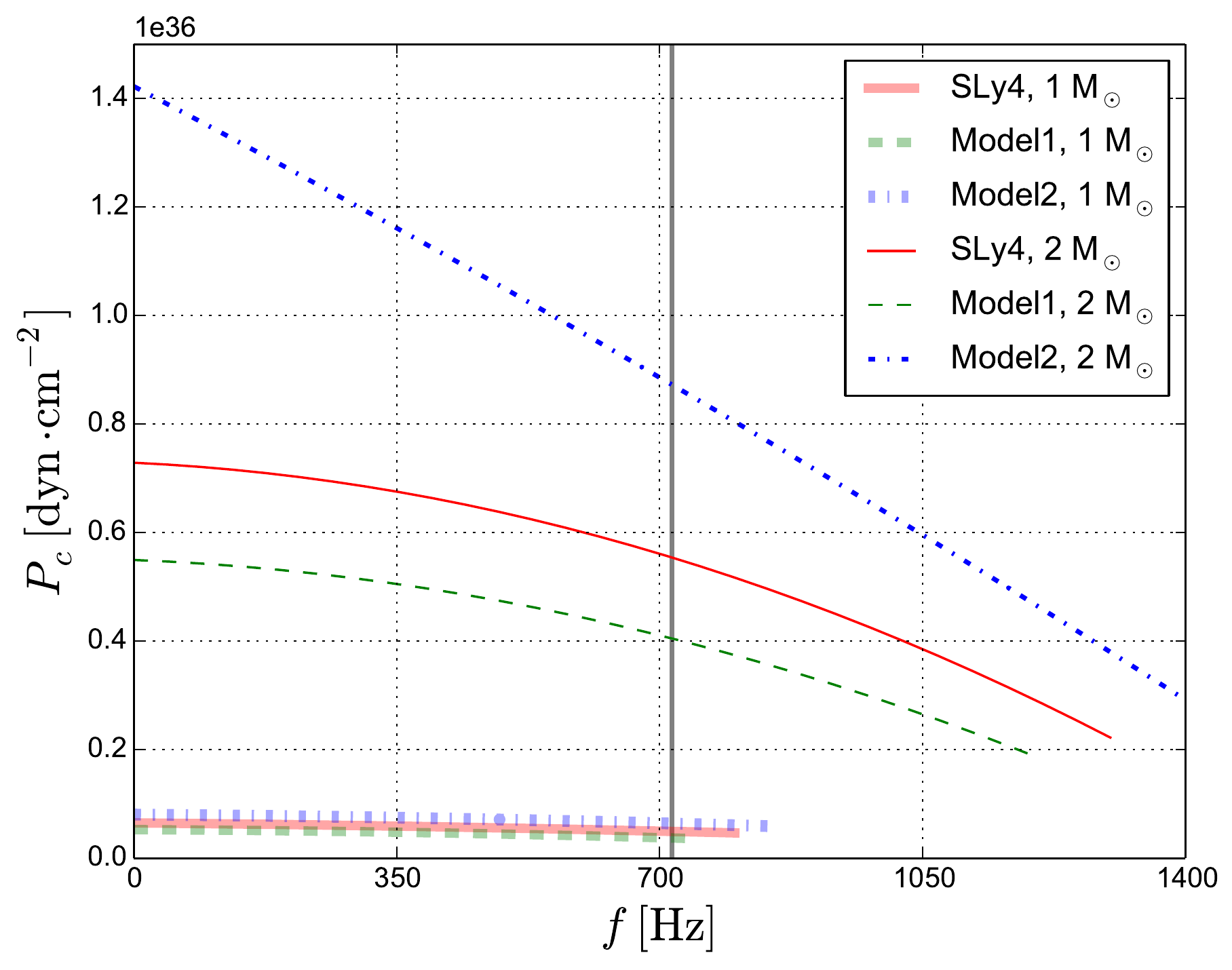}}
\caption{(Color online) Central pressure versus rotational frequency for the SLy4 model (solid red line), Model1 (dashed green line) and Model2 (dashed-dotted blue line), for $1 M_{\odot}$ and $2 M_{\odot}$ masses.} 
\label{fig:pressure}
\end{figure}

We also investigate central EOS parameters (pressure $P_c$, density $\rho_c$
and the baryon density $n_c$) for various masses, spin frequencies and EOS
models. Their behaviour as a function of $f$ strongly depends on the $M$: for
low-mass stars central EOS parameters are almost the same for a broad range of
$f$ (Fig.~\ref{fig:pressure}).
For $M = 1 M_{\odot}$, for all EOSs and $f$, $P_c \approx 7\cdot 10^{35}$
dyn$\cdot$cm$^{-2}$ with negligible errors, $\rho_c \approx 0.75 \cdot 10^{15}$
g$\cdot$cm$^{-3}$ $\pm10\%$ and $n_c \approx 4.1 \cdot 0.1$ fm$^{-3}$ $\pm
15\%$. For NSs with masses $\approx 2 M_{\odot}$, above few hundreds Hz,
the decrease in the values  of central parameters, with increasing rotational
frequency, is rather fast especially for Model2. For example, the difference in
$P_{c}$ between $0$ Hz and $1400$ Hz configurations is almost five times smaller (Fig.~\ref{fig:pressure}). This case, once again,
shows how important the knowledge of $f$ is. Large uncertainties exist also
between the interiors of various models: the difference between SLy4 and Model2
is $\sim 40\%$ in $P_c$ and $\sim 35\%$ in $\rho_{c}$ and $n_{c}$. The difference
between SLy4 and Model1 is slightly less: $\sim 27\%$ in $P_c$ and $\sim
20\%$ in $\rho_{c}$ and $n_{c}$. Compared to global parameters
$\mathcal{O}$ and $\mathcal{S}$, estimates of $P_c$, $\rho_c$ and $n_c$
result in larger differences. For larger spin frequencies, the central
parameters converge to similar values.  

\begin{figure}[ht]
\resizebox{\columnwidth}{!}
{\includegraphics{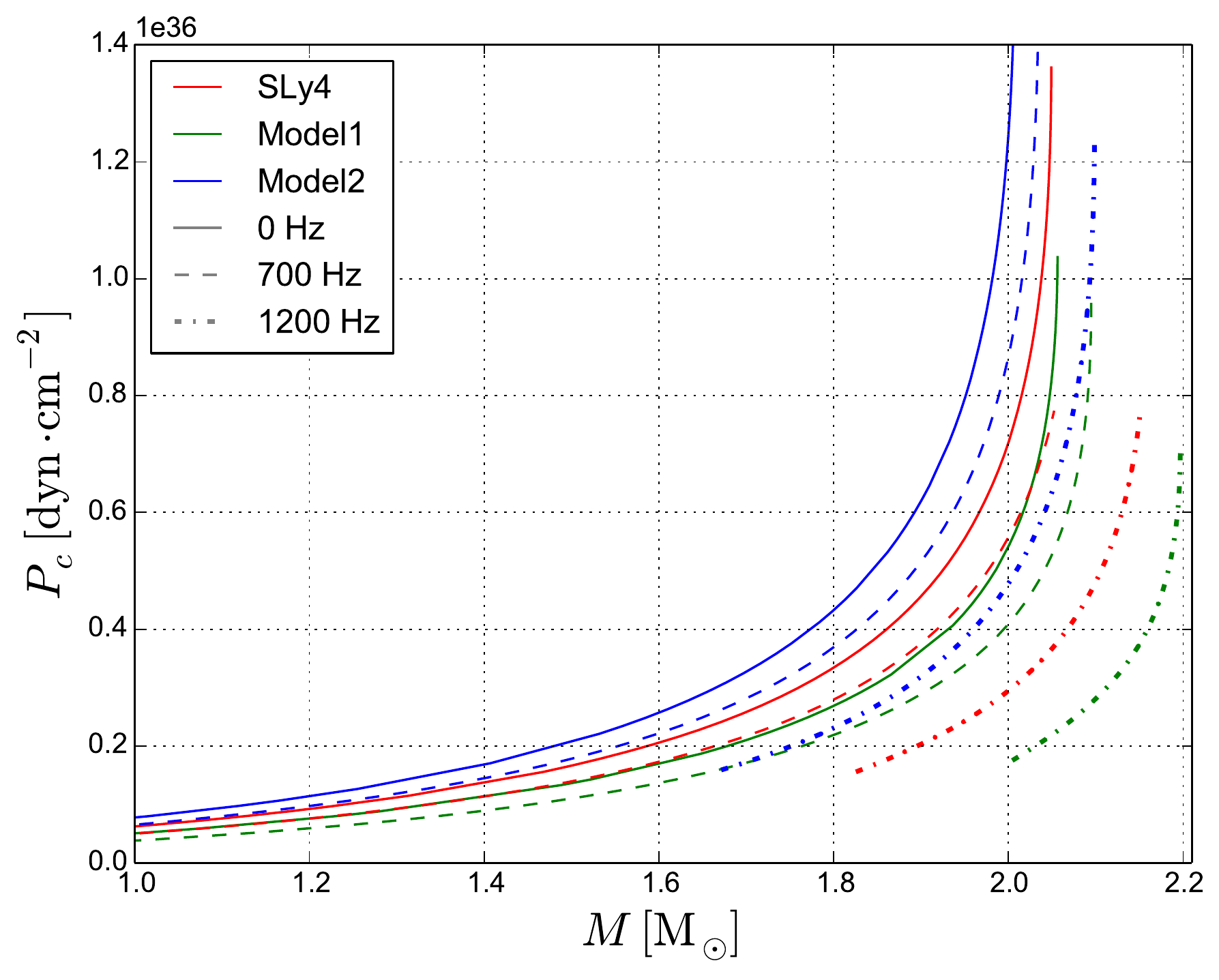}}
\caption{(Color online) Central pressure as a function of mass for the SLy4 model (red), Model1 (green), Model2 (blue), for rotational frequencies 0 Hz (solid), 700 Hz (dashed) and 1200 Hz (dashed-dotted).} 
\label{fig:pressureM}
\end{figure}

\subsection{Known rotational frequency}
\label{sect:knownrotNS}

In Fig.~\ref{fig:pressureM}, the relation between central pressure
$P_{c}$ and mass $M$ is shown for different spin frequencies.
$P_{c}$ slowly increases up to the mass ${\approx}1.9 M_{\odot}$, whereas for
higher $M$ the growth of $P_{c}(M)$ is rapid. This is reflected in the
uncertainty of the central parameters: for $M \lesssim 1.9 M_{\odot}$,
differences between models for $P_{c}$ and $\rho_{c}$, are $\lesssim 10\%$
(this value increases up to $30\%$ when $f$ is unknown). For objects with
higher masses ($\gtrsim 1.9 M_{\odot}$), uncertainties in $P_{c}$ are around
$50\%$, if $f$ is known, and increase to $85\%$ if
$f$ is unknown. Increasing uncertainty in central parameters
at higher masses is related to the softening of the $M(P_c)$ curve due to 
general-relativistic effects near the maximum mass. Although massive stars are much 
more interesting from the point of view of the dense-matter EOS, they are more 
challenging to study than low-mass NSs.  

\begin{figure}[ht]
\resizebox{\columnwidth}{!}
{\includegraphics{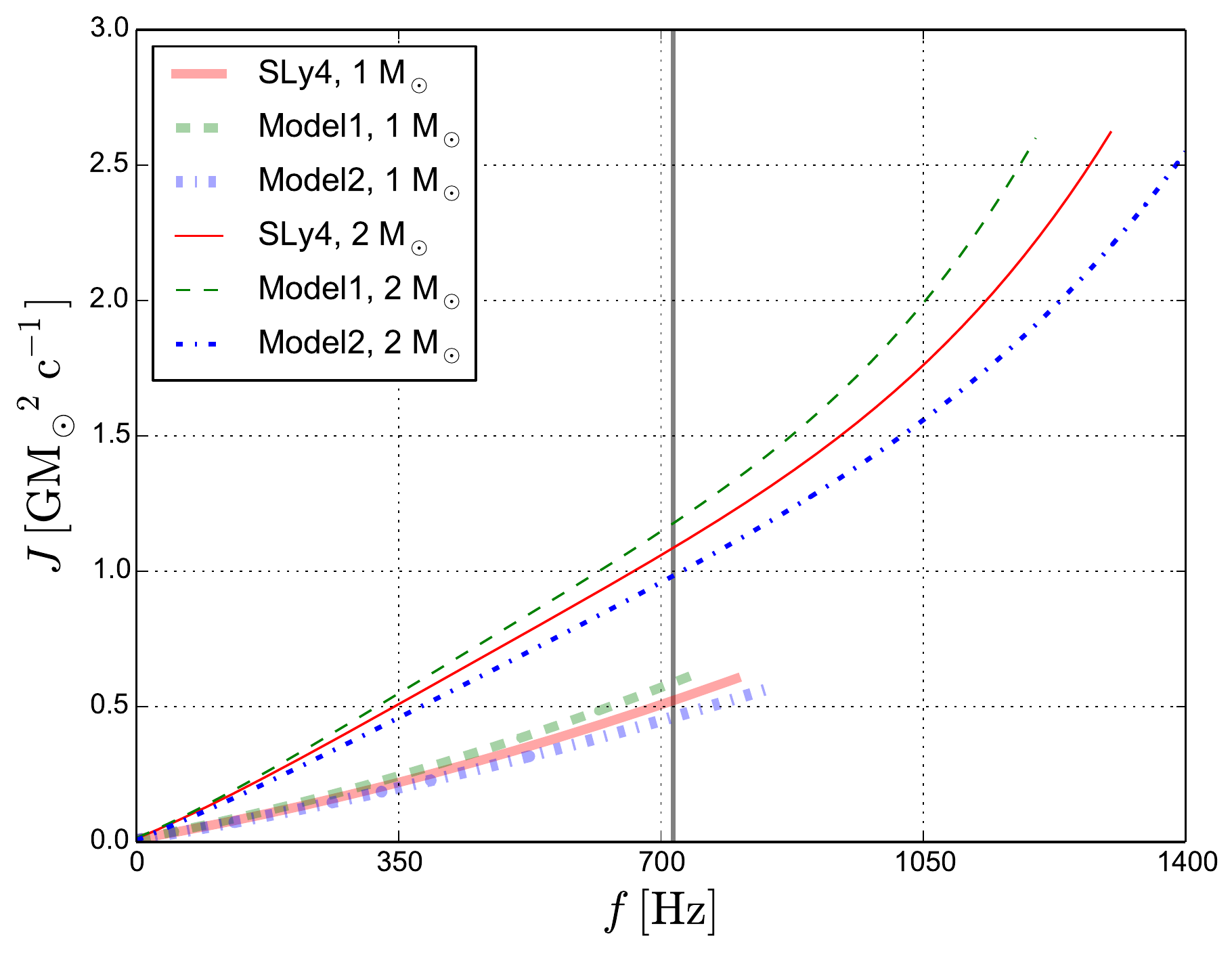}}
\caption{(Color online) Angular momentum versus rotational frequency 
for the SLy4 model (solid red line), Model1 (dashed green line) and Model2 (dash-dotted blue line), for $1 M_{\odot}$ and $2 M_{\odot}$ masses.} 
\label{fig:j}
\end{figure}

In Fig \;\ref{fig:j}, the global angular momentum $J$ for NSs with $M = 1
M_{\odot}$ and $2 M_{\odot}$ is shown. For a specified mass, results for all
three models are very similar for frequencies $\lesssim 1000$ Hz. Above this point one can observe a fast increase in $J$ until the object reaches the
Keplerian frequency. Differences between models for low-mass and
slowly-rotating massive NS are much smaller than for rapidly-rotating massive
stars. The largest differences, about $\pm 20\%$, are present for
sub-millisecond rotation rates. 

\begin{figure}[ht]
\resizebox{\columnwidth}{!}
{\includegraphics{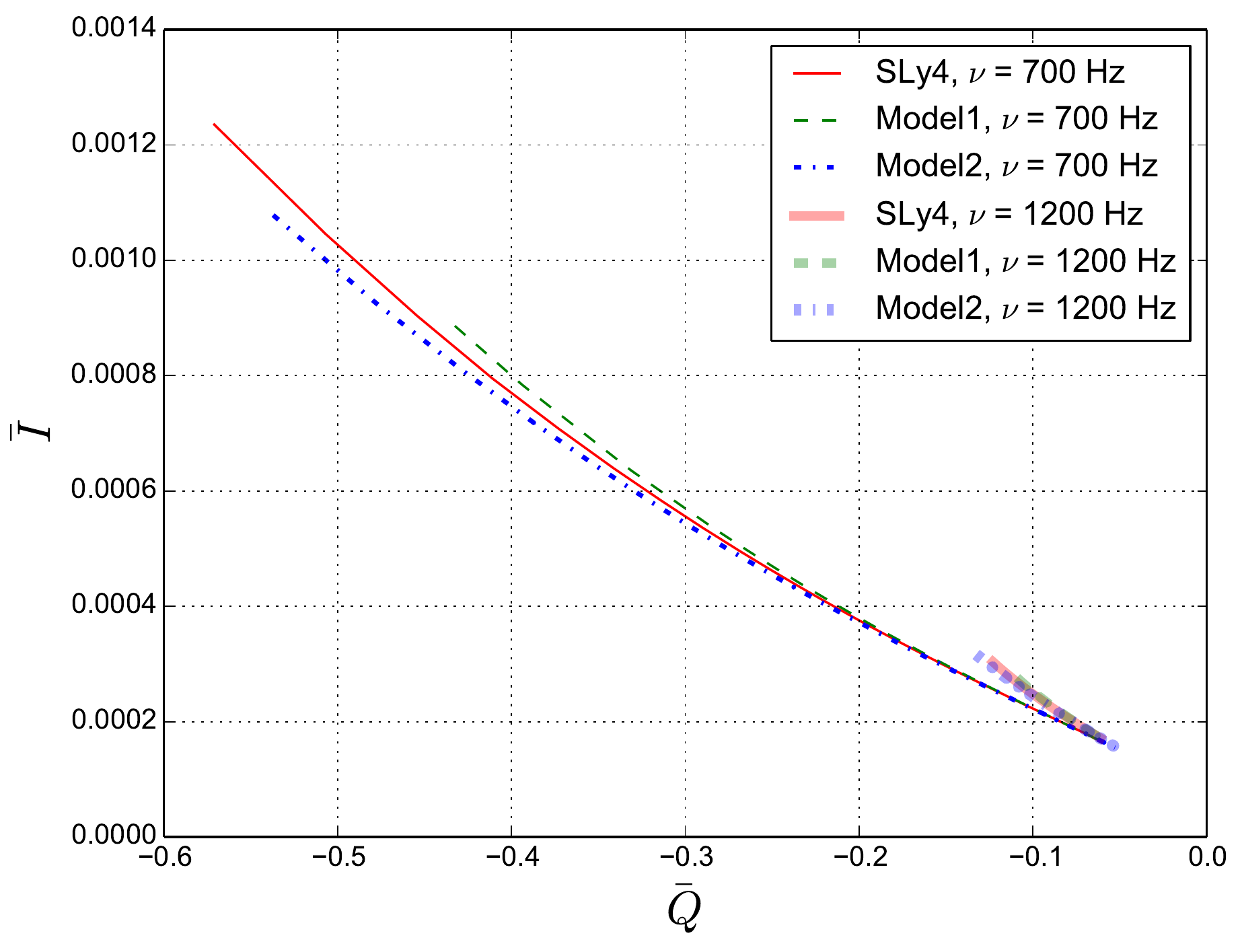}}
\caption{(Colour online) Normalized moment of inertia $\bar{I}$ versus normalized quadrupole moment $\bar{Q}$  
for the SLy4 model (solid red line), Model1 (dashed green line) and Model2 (dash-dotted blue line), 
for $700$ Hz and $1200$ Hz spin frequencies.}
\label{fig:iq}
\end{figure}

According to \citet{Yagi2013} there exists a universal (independent on the EOS)
relation between the quadrupole moment $Q$ and the moment of inertia $I$. It might be
used, for example, to determine rotational frequencies of NSs or distinguish
between ''normal'' NSs and strange stars (see e.g., \citealt{Urbanec2013,Yagi2013}), 
or employed in the description of binary NS inspiral waveforms. We use the following normalization: $\bar{I} = I/M^{3}$ and $\bar{Q} =
Q/(M^{3}\chi^{2})$, where $M$ is gravitational mass, $\chi = J/M^{2}$ and
$J$ is angular momentum of the object. Our results presented in Fig.~\ref{fig:iq} 
show the $\bar{I} - \bar{Q}$ relation for the three models, for two different spin 
frequencies: $700$ and $1200$ Hz. The results are consistent with the 
$I$-$Q$ relation. This is especially true for massive stars (occupying the lower right corner 
of Fig.~\ref{fig:iq}), which are less deformed by the centrifugal force. 

\subsection{$\pm 10\%$ uncertainty in radius measurements}

As mentioned in Sect.~\ref{sect:intro}, $\pm 5\%$ errors in $R$ may be
considered too optimistic and a $\pm 10\%$ uncertainty is a more realistic value
\citep{Lo2013, Miller2016}. In order to study this case two additional piecewise 
polytropic EOSs with the SLy4 EOS crust were constructed. The parameters of the Model3 
and Model4 EOSs are collected in Table~\ref{tab:poly_parameters2}.
\begin{table}[h]
\centering
\begin{tabular}{|c|c|c|}
\hline
par. & Model3 & Model4 \\
\hline
$\gamma_{1}$ & 3.20 & 2.20 \\
\hline
$\gamma_{2}$ & 2.89 & 3.45 \\
\hline
$\gamma_{3}$ & 2.55 & 3.25 \\ 
\hline
$n_{b, 12}$ & 0.14 & 0.25 \\
\hline
$n_{b, 23}$ & 0.37 & 0.35 \\
\hline
$m_{b,1}$ & 1.017369 & 1.015574 \\
\hline
$m_{b,2}$ & 1.016012 & 1.023991 \\
\hline
$m_{b,3}$ & 1.002758 & 1.022353 \\
\hline
$\kappa_{1}$ & 0.008679 & 0.006593 \\
\hline
$\kappa_{2}$ & 0.009633 & 0.002100 \\
\hline
$\kappa_{3}$ & 0.01503 & 0.002694 \\
\hline

\end{tabular}
\vskip 0.5em 
\caption{Parameters of the three polytropes employed for the EOSs of Model3 and
Model4 (indices correspond to the number of the polytrope). }
\label{tab:poly_parameters2}
\end{table}

Model3 and Model4 were chosen to reproduce $\Delta R = \pm10\%$ errors of the radius measurements of the non-rotating reference SLy4 model. Results, as before, depend on the NS mass. Trends are similar to the $\pm 5\%$ case (NSs with masses $\approx 1M_{\odot}$ are more favourable for the estimations of the central parameters like $P_c, \rho_c, n_c$, whereas massive stars produce small errors in the determination of global parameters like $\mathcal{O}$ and $\mathcal{S}$). As expected, most of the uncertainties increase with increasing $\Delta R$.

For low-mass NSs, the accuracy of the $\mathcal{O}$ determination is comparable with the result for the $\Delta R=\pm5\%$ assumption (the oblateness for NSs with masses $\approx 1M_{\odot}$ depends weakly on the accuracy of the $R$ measurements). For massive NSs, the $\mathcal{O}$ uncertainty is $\approx 3\%$ for the frequency of the fastest-known pulsar and $\approx 18\%$ for $1200$ Hz (for the $\pm 5\%$ case the values were $1\%$ and $11\%$, respectively). The errors in the estimation of the surface area $\mathcal{S}$ do not depend on frequency: for the whole spin range they are similar and around $22\%$ (twice what they are in the case of $\Delta R = \pm 5\%$). 

\begin{figure}[ht]
\resizebox{\columnwidth}{!}
{\includegraphics{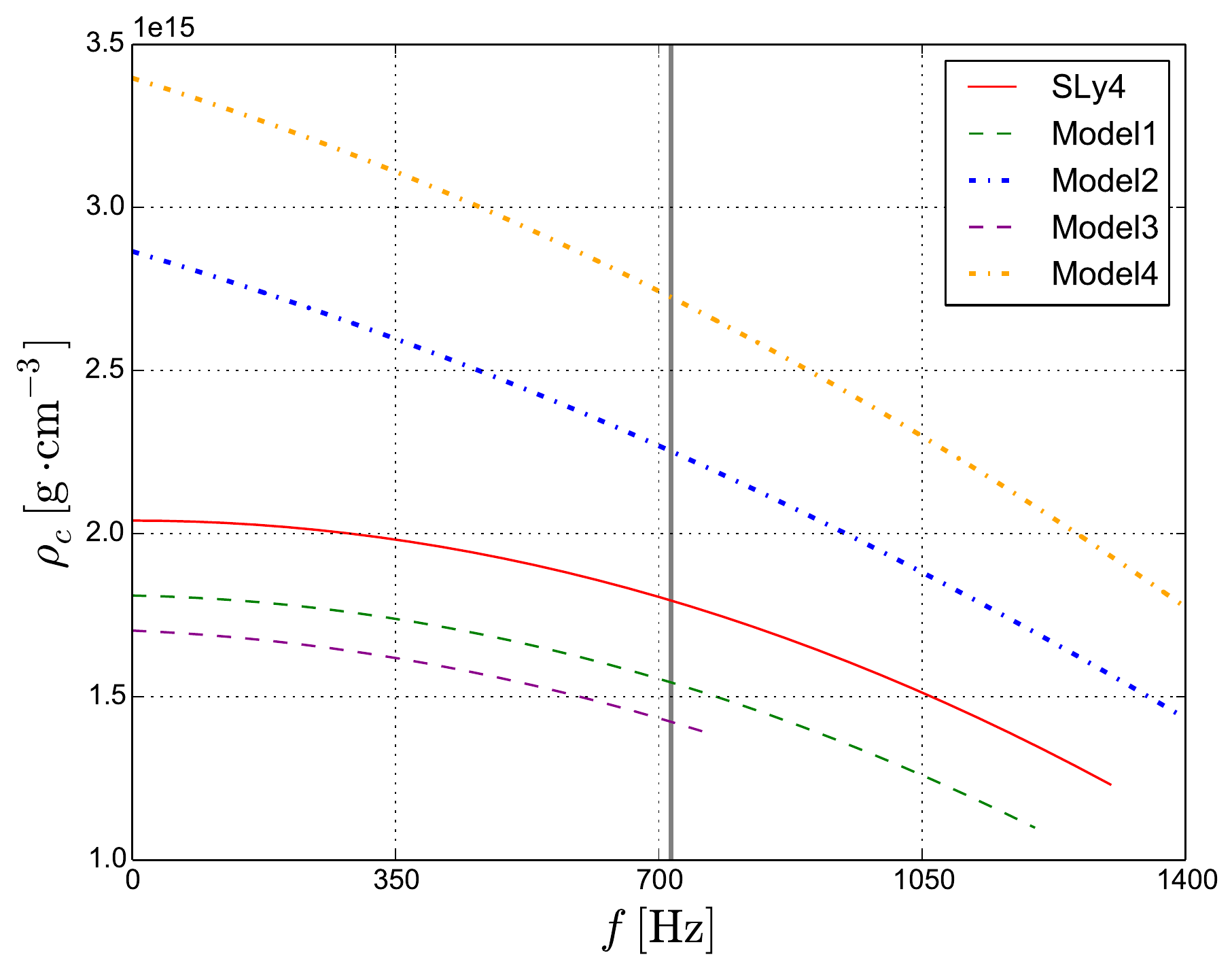}}
\caption{(Color online) Central density $\rho_c$ versus rotational frequency for the SLy4 EOS (red), Model1 EOS (green), Model2 EOS (blue), Model3 EOS (purple) and Model4 EOS (orange), for $2M_{\odot}$.}
\label{fig:cent10}
\end{figure}

As expected, uncertainties in the central parameters are smaller for low-mass stars. For NSs with $M\approx1 M_{\odot}$: $P_c \approx 7\cdot 10^{35}$ dyn$\cdot$cm$^{-2}$ with negligible errors (similarly as for $\Delta R = \pm 5\%$), $\rho_c \approx 0.75 \cdot 10^{15}$ g$\cdot$cm$^{-3}$ $\pm30\%$ ($10\%$ in case $\Delta R = \pm 5\%$) and $n_c \approx 4.1 \cdot 0.1$ fm$^{-3}$ $\pm 30\%$ ($15\%$ in case $\Delta R = \pm 5\%$). For massive stars again one can observe a very fast decrease in the central values with frequency, especially for Model4. An example of the central density $\rho_c$ as a function of the rotational frequency, for all five models $2M_{\odot}$ NS is shown on Fig.~\ref{fig:cent10}. Differences between SLy4 and Model4 are as follows: $64\%$ in $P_c$, $38\%$ in $\rho_c$ and $32\%$ in $n_c$, and between SLy4 and Model3 are $33\%$ in $P_c$, $19\%$ in $\rho_c$ and $15\%$ in $n_c$.

\subsection{Tidal deformability}

\begin{figure}[ht]
\resizebox{\columnwidth}{!}
{\includegraphics{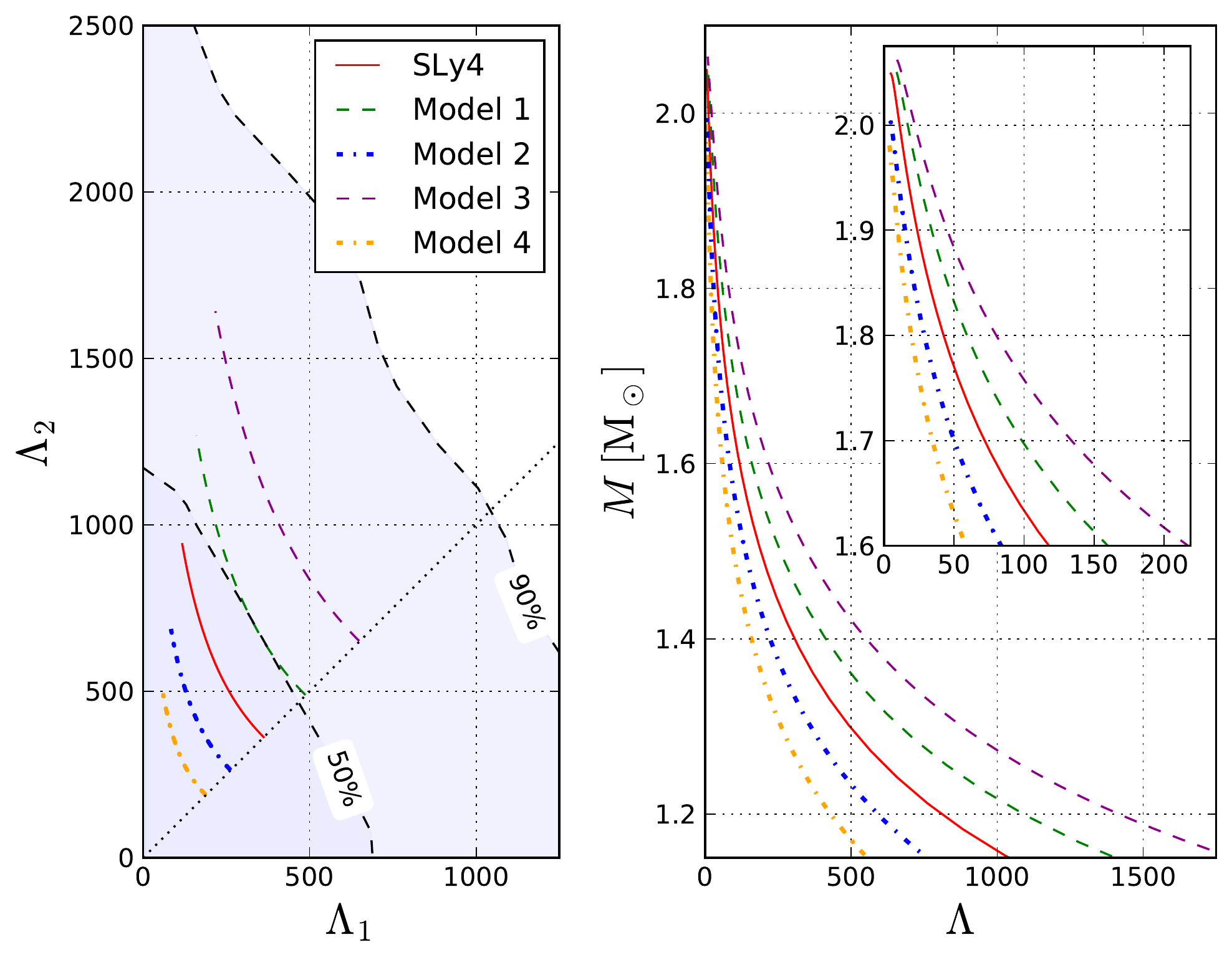}} 
\resizebox{\columnwidth}{!}
{\includegraphics{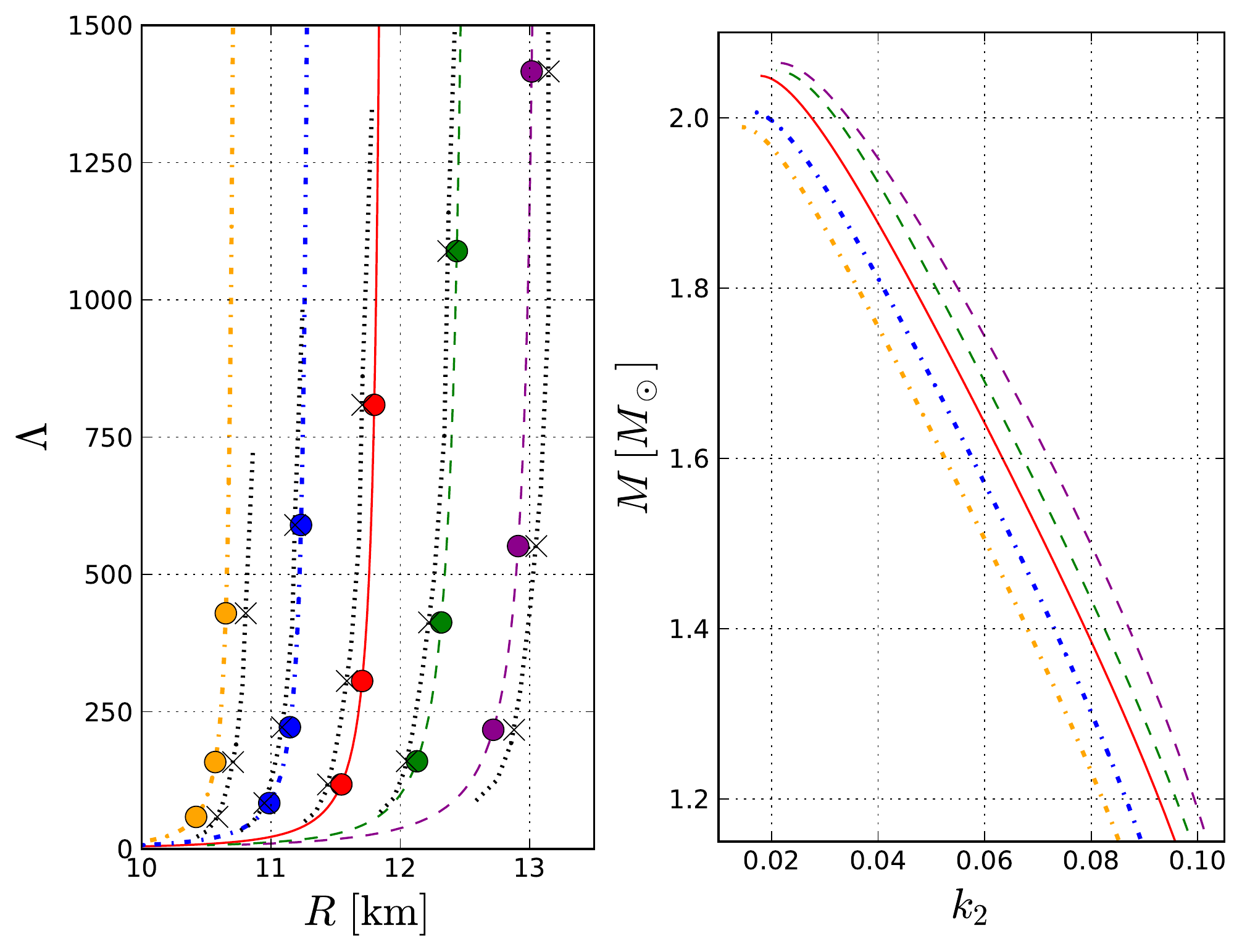}}
\caption{(Color online) Upper row, left panel: Comparison of tidal deformabilities $\Lambda$ for stellar component masses compatible with the GW170817 observation (assuming low-spin priors, \citealt{Abbott2017a}). The shaded area denotes the estimated 90\% confidence region corresponding to the measurement. Right panel: Stellar mass as a function of tidal deformability $\Lambda$, for the mass range between 1.15 $M_\odot$ and the $M_{max}$. Lower row, left panel: tidal deformability $\Lambda$ as a function of stellar radius; points mark 1.2, 1.4 and 1.6 $M_\odot$ (from top to bottom). Approximate formula denoted in the text is represented by black dotted curves. Right panel: mass as a function of the $k_2$ tidal Love number.} 
\label{fig:l1l2-ml}
\end{figure}

We also investigate the recent gravitational-wave estimate of the tidal
deformabilities in the GW170817 binary NS system. Figure~\ref{fig:l1l2-ml}
reproduces the $\Lambda_1 - \Lambda_2$ relation for the deformabilities in the
range of masses corresponding to the measured chirp mass $\mathcal{M} =
\left(M_1 M_2\right)^{3/5}/\left(M_1 + M_2\right)^{1/5} =
1.188^{+0.004}_{-0.002}\ M_\odot$ and the range of component masses in the case
of low-spin priors, $M_1 = 1.36 - 1.60\ M_\odot$ and $M_2 = 1.17 - 1.36\
M_\odot$ (the definition of the tidal deformability $\Lambda$ was introduced in
Sect.~\ref{sect:methods}).  For the three EOS models discussed here, the
difference in radius $\Delta R=\pm5\%$ is consistently reflected in the
$\Delta\Lambda$ values: between 250 for $1.17\ M_\odot$, and 50 for $1.6\
M_\odot$. Increasing $\Delta R$ by a factor of two gives a two
times larger $\Delta\Lambda$. This trend can be understood by analysing
Fig.~\ref{fig:profiles}, where the $P(\rho)$ ranges for GW170817 component
masses are displayed, as well as Fig.~\ref{fig:l1l2-ml}, where the relation
between $R$, $M$, $\Lambda$ and $k_2$ are plotted. The EOSs yield, to first
approximation, a constant difference in the stellar radius, and have similar
stiffness in the relevant mass ranges, which is reflected in the $k_2$ values,
while the difference in $\Lambda$ between models is dominated by the $M^{-5}$
term. Regular behaviour of $\Lambda$ with $R$ and $M$ may be exploited to
approximately reproduce $R$, given $\Lambda$ and $M$ values. A linear relation
$R\approx a(M)\Lambda + b(M)$ [km], where $a(M) = \sum_{i=0}^4 a_iM^i$, $b(M) =
\sum_{i=0}^1 b_iM^i$ ($a_4 = 0.1332$, $a_3 = -0.6426$, $a_2 = 1.1886$, $a_1 = -0.9855$, 
$a_0 = 0.3076$, $b_1 = -0.1133$, $b_0 = 9.9216$, $M$ is the $M_\odot$ units) recovers 
the values of radii $R$ for the models considered here with an error of typically 
0.1 km. 

\section{Discussion}
\label{sect:discussion}

In this work we have estimated how much information about the EOS parameters
can be drawn from current and future measurements of neutron star radii,
assuming a target accuracy of about $5\%$. To address this question, we have
compared the widely accepted SLy4 EOS \citep{Douchin2001}, treated here as a
reference EOS, with two parametric EOSs designed to yield TOV $M(R)$ sequences
with $\pm 5\%$ of the SLy4 radius.  

We have also considered the influence of rigid axisymmetric rotation on the
global properties of the NSs and their central EOS parameters. In some cases,
certain configurations may mimic other ones with different rotational
frequencies and EOSs. We show that even if $M$ is established precisely and errors are present only in the measurement of $R$, lack of the information about
$f$ might lead to $40\%$ error in central pressure estimation (see Figs
\ref{fig:MR} and \ref{fig:MRspecial_pres}). 

Rapid rotation may be a crucial factor in the distinction between NSs with
different EOSs. For spins comparable with the rotational frequency of the
fastest known pulsar (716 Hz for PSR J1748-2446ad), the radii deviate by more
than $5\%$ from the reference SLy4 radius for stars with masses $\lesssim
1.4M_{\odot}$, whereas for $f = 1200$ Hz the radius difference is larger than
$5\%$ for the whole available NS mass range.  As expected, for low-mass stars,
rotation manifests itself strongly by deforming the star, but the central
parameters are almost unchanged.  A $5\%$ accuracy in the measurement of $R$
leads to $8-10\%$ errors in the estimation of the stellar
oblateness $\mathcal{O}$ and the surface area $\mathcal{S}$.  Even if the
rotational frequency of these stars is unknown, their central properties can be
established with $\pm 10\%$ accuracy. For more massive stars, with masses
$\approx 2M_{\odot}$ and known rotational frequency $f$, the uncertainties are
much higher, up to $40\%$.

For cases when only compactness can be measured (e.g., because of the slow
rotation of the object), it is possible to estimate surface area with $\sim
10\%$ accuracy, for sources with known spin frequency. 

We take as an example PSR J1748-2446ad, and show how limited our reconstruction of
the properties of a compact object is, when only its rotational frequency is
known. Assuming, for the sake of an example, that the SLy4 EOS is the true EOS,
one can expect that the oblateness lies between $0.72$ and $0.97$ for masses
between $1-2 M_{\odot}$. If we assume $\pm 5 \%$ observational uncertainty on
radius, it leads to an $\approx 30\%$ accuracy in the determination of $\rho_{c}$,
$\approx 20\%$ for $n_{b}$ and $\approx 50 \%$ accuracy for $P_{c}$, if this
object is massive. If its mass is only $\approx 1M_{\odot}$, central parameters
for various models are very similar to each other. 

We repeated our study for the $\Delta R = 10\%$ case. As expected,
uncertainties on the central and global parameters increase. Oblateness
$\mathcal{O}$ can be estimated with $\lesssim 18\%$ and $\mathcal{S}$ with
$\approx 22\%$ error, which are approximately twice the measurment uncertainties of the $\pm5\%$ case. 
 The accuracy of the estimation of the central
parameters depends on the NS mass. For low-mass stars $P_c$ can be estimate with
negligible errors, and $\rho_c$ and $n_c$ with $\approx 30\%$ uncertainties. For
massive NSs errors are up to $64\%$ in $P_c$, $38\%$ in $\rho_c$ and $32\%$ in
$n_c$. These results suggest that increasing $\Delta R$ by a factor of two decreases
the  accuracy of the estimation of the NS central 
parameters by a factor of ${\sim}2$ to $3$.

We also estimate the effect of the radius uncertainty on the tidal
deformability $\Lambda$, using as an example the estimated masses of the
components from the GW170817 event \citep{Abbott2017a}. The $\pm 5\%$ radius
difference of the parametric models with respect to the SLy4 model results in
$\Delta\Lambda$ between 250 at $1.17\ M_\odot$ and 50 at $1.6\ M_\odot$ 
( these values are two times higher for $\Delta R=\pm10\%$), and
since the EOSs possess similar stiffness in the relevant regimes and hence
similar behavior of the tidal Love numbers $k_2$ (Figs.~\ref{fig:profiles} and
\ref{fig:l1l2-ml}), the $\Delta\Lambda$ is mostly influenced by the dependence
of $\Lambda$ on mass $M$.  

Linking the NS global parameters with the properties of the dense matter EOS 
is an area of active effort in view of forthcoming observational results. 
An extensive study on the full currently allowed range of NS radii will be 
the subject of future work. 


\begin{acknowledgements}

We acknowledge Prof. J. L. Zdunik for his implementation of the tidal Love
numbers in the TOV equation and Prof. T. G{\"u}ver for useful comments and
suggestions. MS and MB were partially supported by the Polish NCN grants no.
2014/14/M/ST9/00707 and 2016/22/E/ST9/00037. BH and MB acknowledges support from the
European Union’s Horizon 2020 research and innovation programme under grant
agreements No. 70271 and No. 653477, respectively. 
We acknowledge partial support by the `PHAROS' COST action
CA 16214. This research was completed using free and open software ({\tt
LORENE}, {\tt gnuplot}, {\tt python}: {\tt matplotlib}, {\tt numpy} and {\tt
scipy}). 

\end{acknowledgements}


\end{document}